\documentclass[prb,aps,twocolumn,amsmath,amssymb,floatfix,
superscriptaddress, raggedbottom]{revtex4-2}
\usepackage[dvips]{graphics}
\usepackage[justification=raggedright]{caption}
\usepackage{amsmath}
\usepackage{subcaption}
\usepackage{physics}
\usepackage{color}
\usepackage{float}
\usepackage{natbib}
\usepackage[dvipsnames]{xcolor}
\setcitestyle{numbers}
\setcitestyle{square}
\definecolor{dred}{rgb}{0,0,0.6}
\usepackage{graphicx}
\usepackage{bm}  
\usepackage{amssymb}  
\usepackage{amsmath}
\usepackage{braket}
\usepackage[mathscr]{euscript}
\usepackage[colorlinks=true, allcolors=blue]{hyperref}
\begin{document}
\title{Wannier charge center, spin resolved bulk polarization and corner modes in a strained quantum spin Hall insulator}

\author{Srijata Lahiri and Saurabh Basu \\ \textit{Department of Physics, Indian Institute of Technology Guwahati-Guwahati, 781039 Assam, India}}
\date{\today}
\begin{abstract}
{Topological invariants are a significant ingredient in the study of topological phases of matter that intertwines the supposedly contradicting concepts of bulk and boundary. The nature of the invariants differ depending on the dimensionality of the boundary at which the topologically non-trivial states manifest themselves. The primary motivation of this work is to study two distinct scenarios of topological phase, differing in the dimensionality of their boundary states and study the associated bulk topological invariants that characterize them. In this regard, we study the band engineered Kane Mele model which originally is a prototypical example of a system that hosts quantum spin Hall effect on a honeycomb lattice.  Under a smooth band deformation caused by varying one of the nearest neighbor hopping amplitudes (say $t_1$) as compared to the other two (say $t$), we observe that the system transits from its first order topological insulating state (or quantum spin Hall state) to a second order topological insulating (SOTI) state via a gap closing transition. This transition occurs when the system crosses a particular threshold of the deformation parameter $t_1\mathbin{/}t$ (namely $t_1\mathbin{/}t=2$), known as the semi-Dirac limit in literature. We show the presence of edge and corner modes as a signature of first and second order topology respectively. Further, we observe the evolution of the Wannier charge center (WCC), a bulk property as a function of the deformation parameter ${t_1}\mathbin{/}{t}$. It is seen that the behavior of the WCC is entirely different in the quantum spin Hall (QSH) phase as compared to the second order topological state. We also find that, while the $\mathbb{Z}_2$ invariant successfully characterizes the QSH state, it cannot characterize higher order topology (second order here). The model being mirror invariant, we also calculate mirror winding number to show that it is rendered trivial in the SOTI phase as well, while being non-trivial in the QSH phase. Finally, we observe that the spin resolved bulk polarization correctly establishes the appearance of second order topological corner modes and thus categorizes this phase as an \textit{obstructed atomic insulator}.  }

\end{abstract}
\maketitle
\section{\label{sec:level1}Introduction}
Topological insulators (TI) have been a subject of extensive research in the past decade. TIs are novel materials that show the intriguing feature of hosting a gapped bulk, but gapless edge/surface states. These topological states are robust and protected against minor perturbations that do not disturb the symmetries inherent in the system. Traditional TIs show the essence of non-trivial topology on a $d-1$ dimensional surface for a bulk that is $d$-dimensional \cite{Murakami_2011,RevModPhys.82.3045, PhysRevLett.118.076803, Anomalous_QHE}. A major aspect of topological materials lies in the bulk boundary correspondence, where a topological invariant, evaluated purely from the bulk eigenstates, predicts the behaviour at the boundaries of the system. Currently there have been multiple extensions to the field of topological materials. These include floquet topological insulators which exhibit topological phases exclusive to a periodically driven system and not shown by their static counterpart \cite{FTI1,FTI2,FTI3,FTI4}. Furthermore, non-Hermitian (NH) TIs are gaining growing attention recently \cite{NH1, NH2, NH3, NH4, NH5}. Non-Hermiticity enhances the richness of topological phases of matter and cannot be bound within the conventional \textit{10-fold classification} of symmetry protected topological states. NH systems also feature topological manifestations unusual to their Hermitian counterparts which include the presence of exceptional points and skin effect. Another important extension to the field of topological insulators, that is being actively explored, are higher order topological insulators (HOTI) \cite{PhysRevResearch.3.L042044,PhysRevLett.123.256402,PhysRevB.96.245115,PhysRevLett.119.246402,PhysRevB.97.155305,PhysRevLett.120.026801,Costa2021,Noguchi2021,PhysRevB.101.235403,PhysRevB.106.205111,PhysRevLett.121.075502, PhysRevD.13.3398,PhysRevLett.106.106802, Arnob1, Arnob2, Anomalous_HOTI, Anomalous_floquet, Quasicrystal_HOTI, Graphene_floquet_HOTI}. Unlike conventional TI, an $n^{th}$ order HOTI exhibits the presence of non-trivial topological states on a surface/edge of dimension $d-n$ for a bulk that is $d$ dimensional. This gives rise to corner modes in 2D and corner/hinge modes in 3D HOTI systems. The conventional definition of bulk boundary correspondence fails here. Rather, HOTI shows a refined bulk boundary correspondence. Higher order topological insulators can be majorly called a `subclass' of topological crystalline insulators where rotation or mirror symmetries protect the topological phases. Research in this field found a massive boost with the advent of the electric multipole insulators in the study of Benalcazar \textit{et al} \cite{Benalcazar} as well as chiral and helical higher order topological states in the study of Schindler \textit{et al} \cite{Schindler}. In the latter work, prospective material candidates such as SnTe and surface modified BiTe and BiSe, have been theoretically claimed to host a higher order phase. However, despite HOTI being a well studied phenomena in recent times, it is still ambiguous how the topological conducting edge/surface states of a 2D/3D system can be gapped out to show higher order topology. {Here, we study one such possibility where band deformation under strain of a quantum spin Hall insulator induces a transition from a TI to an HOTI phase of matter.}\par Our primary aim is to study two topological phases of different order, one evolving into the other and track the behavior of the corresponding bulk topological invariants that characterize them. In this regard, the Kane Mele model which is a prototypical example of the quantum spin Hall insulator, is considered \cite{KM}. The proposal of the Kane Mele model owes its origin to the seminal work by Haldane who showed that an external magnetic field and hence Landau levels are not indispensable for the observation of quantum Hall effect \cite{HM}. Haldane introduced a complex second neighbour hopping to a honeycomb lattice which causes the Dirac nodes at the \textbf{K} and \textbf{K$'$} points in the Brillouin zone (BZ) of bare graphene to gap out, thus giving rise to conducting edge states. This complex second neighbour hopping however breaks time reversal symmetry (TRS) and bestows the occupied energy subspace with a non-zero Chern number, thus yielding a non-zero conductance similar to the original quantum Hall effect. Since the Haldane model breaks TRS, it was now imperative to study how topology behaves if TRS is restored. With this aim, Kane and Mele proposed a spinful model with an equal and opposite Haldane flux for the spin up and spin down particles. The spinful bands acquire opposite Chern number thus causing the net Chern number of the occupied energy subspace to vanish. This is in accordance with TRS. It is however observed that the difference of Chern number for the two spin sectors act as an effective topological invariant implying that the system shows a finite spin Hall conductance although the conductance in the charge sector vanishes. Such systems fall under the class of $\mathbb{Z}_2$ topological insulators and show the presence of helical edge states. Moreover, it was observed by Kane and Mele that an inversion symmetry breaking Rashba spin orbit coupling term, that destroys the conservation of the $z$ component of spin, causes little qualitative difference to the original results owing to leaving the TRS intact. Experimental evidence of the QSH state has been suggested to be found in HgTe/CdTe quantum well \cite{HGTE}, low buckled germanene \cite{LBG}, Cl-doped ZnSe \cite{ZNSE}, Pt wires \cite{PT} etc. It should be mentioned here that the limit $t_1\mathbin{/}t=2$ is largely known as the semi-Dirac limit, where the bulk energy spectrum shows a linear dispersion along one component of momentum and a quadratic dispersion along the direction perpendicular to the former. Evidence of such inhomogeneous dispersion has been expected to be found in monolayer phosphorene subjected to pressure or doping \cite{Semi-Dirac1}, deformed graphene \cite{Semi-Dirac2} etc.
\begin{figure}
    \centering
    \includegraphics[width=\columnwidth]{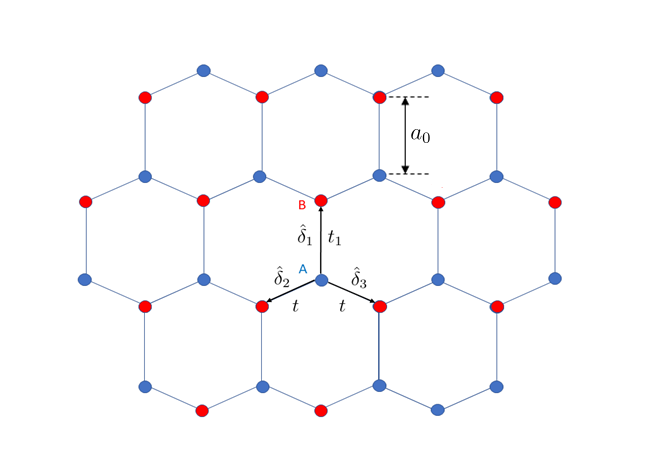}
    \caption{A schematic representation of the honeycomb lattice on which the Kane-Mele model is studied. $\hat\delta_1$, $\hat\delta_2$ and $\hat\delta_3$ represent nearest neighbor vectors. $a_0$ represents the nearest neighbor distance.}
\label{Figk0}
\end{figure}
\par In this work, we smoothly deform the bands of the Kane Mele model defined on a honeycomb lattice by modifying one of the nearest neighbour hopping amplitudes (say $t_1$), while keeping the other two (say $t$) fixed. It is seen that the quantum spin Hall state with distinct edge modes is destroyed beyond the critical point $t_1\mathbin{/}t=2$, and the system converts itself into a second order topological insulator, which is an HOTI with topological states manifested at the $d-2$ dimensional boundary. The bulk bandstructure shows a shift in the band extrema points as a function of the deformation parameter $\xi={t_1}\mathbin{/}{t}$. At $t_1=t$ ($\xi=1$) the bulk bandstructure hosts the band minima at the \textbf{K} and \textbf{K$'$} points in the Brillouin Zone (BZ). They shift towards each other along the $\Gamma-K-M-K'-\Gamma$ line before finally merging at the $M$ point of the BZ when $t_1=2t$ ($\xi=2$). It is seen that the behaviour of the bulk topological invariants corresponding to the two different topological regimes is completely different owing to their dissimilar order. In this regard, we mention another work by Ren \textit{et al} \cite{Zeeman}, where an in-plane Zeeman field applied to the Kane Mele model, destroys the QSH phase and transforms the system into a higher order topological insulator. However, the TRS is broken in this system and the vital essence of the Kane Mele model is lost. On the contrary, in our work we keep the TRS of the system undisturbed while inducing a second order topological phase solely by means of band engineering. While extensive work has been done on several models featuring an HOTI phase, we focus on the transition of the system and the corresponding bulk topological invariants as it smoothly changes its topological order as a function of band deformation. We also provide a clear perspective pertaining to the occurence of this transition which is crucial to the study of topological phases of matter.\par The paper is organized as follows. In section II we define the tight binding Hamiltonian for the strained Kane Mele model and show the effect of band deformation on the bulk bandstructure. The energy spectra of a ribbon-like configuration is also studied which shows the existence of helical edge modes in the regime $\xi<2$. Further deformation destroys the QSH phase and the helical edge modes vanish. However, beyond this critical point, a real space probability distribution shows the existence of zero energy corner modes in the system localized at two corners of a suitably formed supercell that obeys the crystal symmetries of the Hamiltonian. In section III we study the evolution of the Wannier charge center along one direction (say $x$) with respect to momentum along the other (say $y$). It is seen that the nature of this evolution is completely dissimilar for the two different regimes. Correspondingly, the $\mathbb{Z}_2$ invariant which is finite in the region $\xi<2$, vanishes beyond it. Pertaining to the presence of mirror symmetry $M_x$ in the system, we also calculate the mirror winding number which corresponds to the Berry phase picked up by the ground state of a mirror symmetry resolved effective Hamiltonian over a complete cycle in its parameter space. We observe that the mirror winding number shows a similar trend as the WCC. However, the spin resolved bulk polarization which indicates the position of the center of charge in a unit cell becomes quantized in the second order topological phase. This indicates an \textit{obstructed atomic insulator} where the center of charge suffers a mismatch from the original lattice sites \cite{OAI}. This leads to an excess charge accumulation at the corners of a rhombic supercell which manifests as second order topology. Finally we conclude with a brief summary of our results in section IV. 
\section{\label{sec:level2}The Hamiltonian}
The Kane Mele model defined on a honeycomb lattice is shown in Fig. \ref{Figk0}. 
\begin{figure}
\centering
\begin{subfigure}{0.49\columnwidth}
         \centering
         \includegraphics[width=\columnwidth]{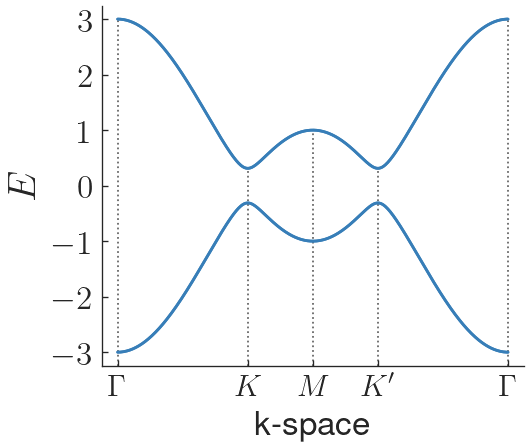}
         \caption{$\xi=1.0$}
         \label{Figk1a}
     \end{subfigure}
\begin{subfigure}{0.49\columnwidth}
         \centering
         \includegraphics[width=\columnwidth]{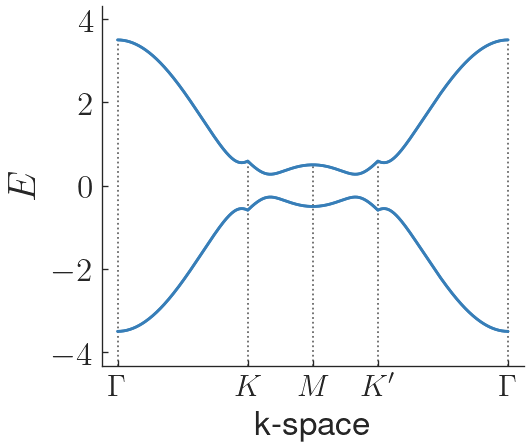}
         \caption{$\xi=1.5$}
         \label{Figk1b}
     \end{subfigure}
\begin{subfigure}{0.49\columnwidth}
         \centering
         \includegraphics[width=\columnwidth]{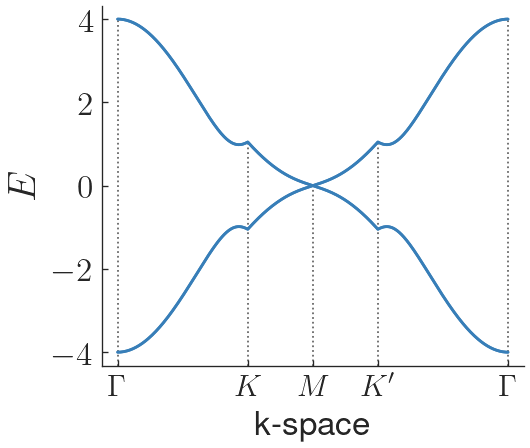}
         \caption{$\xi=2.0$}
         \label{Figk1c}
     \end{subfigure}
\begin{subfigure}{0.49\columnwidth}
         \centering
         \includegraphics[width=\columnwidth]{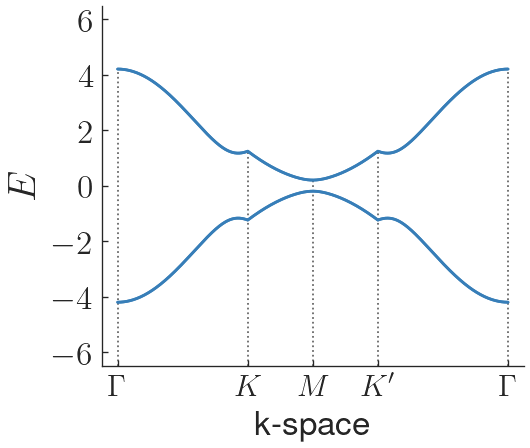}
         \caption{$\xi=2.2$}
         \label{Figk1d}
     \end{subfigure}
\caption{The bulk bandstructure of the band deformed Kane Mele model is shown for different values of the deformation parameter $\xi(=\frac{t_1}{t})$. It is seen that the band extrema shift towards each other as a function of $\xi$, to finally meet at the $M$ point of the BZ for $\xi=2$. As $\xi$ is increased further, the gap reopens, indicating a topological phase transition. The values of $\lambda_{so}$ is fixed at 0.06$t$ while $\lambda_R$ and $\lambda_v$ are kept $0$. This leads to a two-fold degeneracy of the bulk bands.}
\label{Figk1}
\end{figure}
\begin{figure}
\centering
\begin{subfigure}{0.49\columnwidth}
         \centering
         \includegraphics[width=\columnwidth]{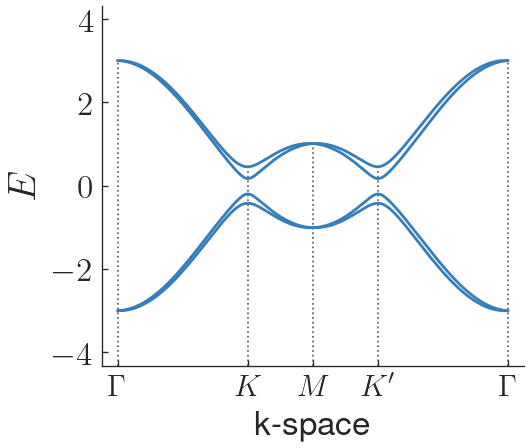}
         \caption{$\xi=1.0$}
         \label{Figk2a}
     \end{subfigure}
\begin{subfigure}{0.49\columnwidth}
         \centering
         \includegraphics[width=\columnwidth]{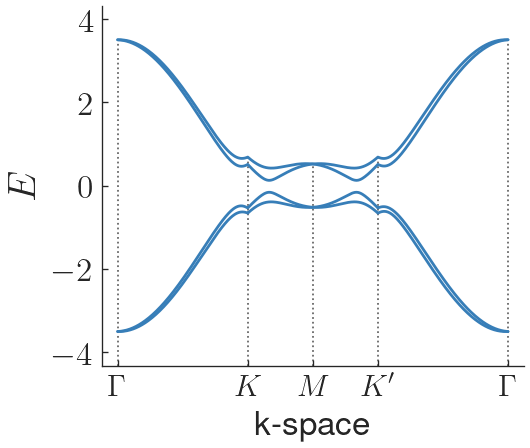}
         \caption{$\xi=1.5$}
         \label{Figk2b}
     \end{subfigure}
\begin{subfigure}{0.49\columnwidth}
         \centering
         \includegraphics[width=\columnwidth]{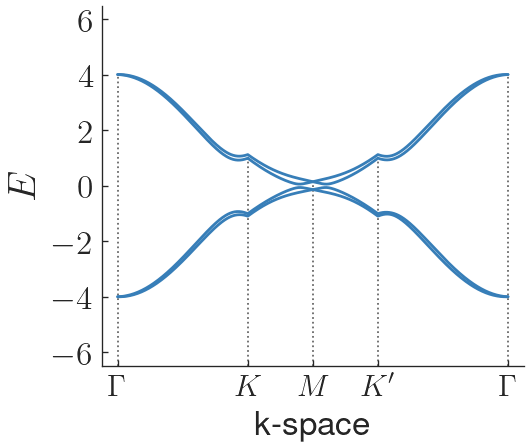}
         \caption{$\xi=2.0$}
         \label{Figk2c}
     \end{subfigure}
\begin{subfigure}{0.49\columnwidth}
         \centering
         \includegraphics[width=\columnwidth]{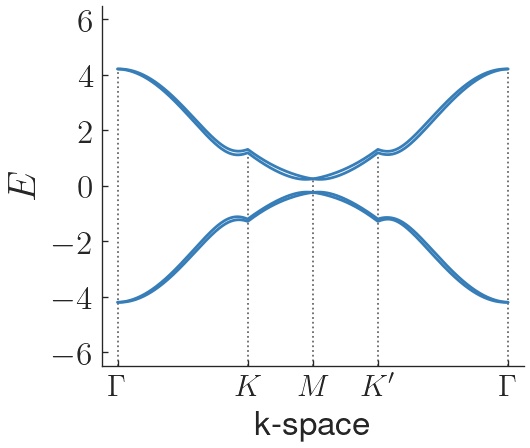}
         \caption{$\xi=2.2$}
         \label{Figk2d}
     \end{subfigure}
\caption{The two-fold degeneracy of the bulk bandstructure is lifted as soon as non-zero values of $\lambda_R$ and $\lambda_v$ are introduced in the model. The behaviour of the bands with respect to the deformation parameter $\xi$ however remains the same as shown in Fig. \ref{Figk1}. Here, values of the parameters are given by $\lambda_{so}=0.06t$, $\lambda_R=0.05t$ and $\lambda_v=0.1t$.}
\label{Figk2}
\end{figure}
\begin{figure}
\centering
\begin{subfigure}{0.49\columnwidth}
         \centering
         \includegraphics[width=\columnwidth]{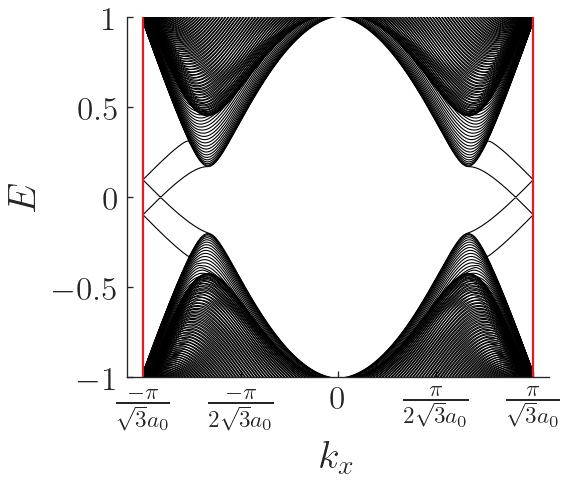}
         \caption{$\xi=1.0$}
         \label{Figk3a}
     \end{subfigure}
\begin{subfigure}{0.49\columnwidth}
         \centering
         \includegraphics[width=\columnwidth]{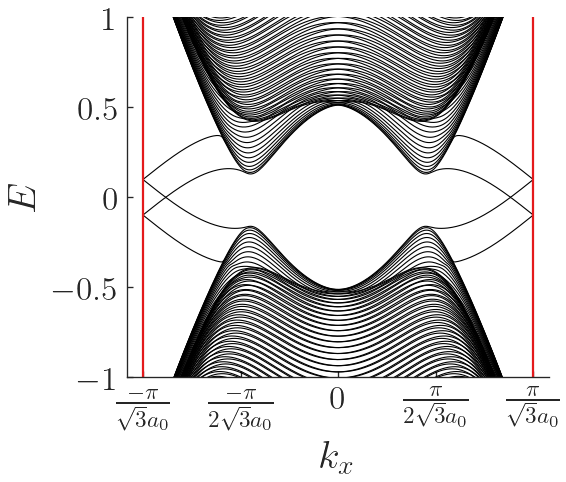}
         \caption{$\xi=1.5$}
         \label{Figk3b}
     \end{subfigure}
\begin{subfigure}{0.49\columnwidth}
         \centering
         \includegraphics[width=\columnwidth]{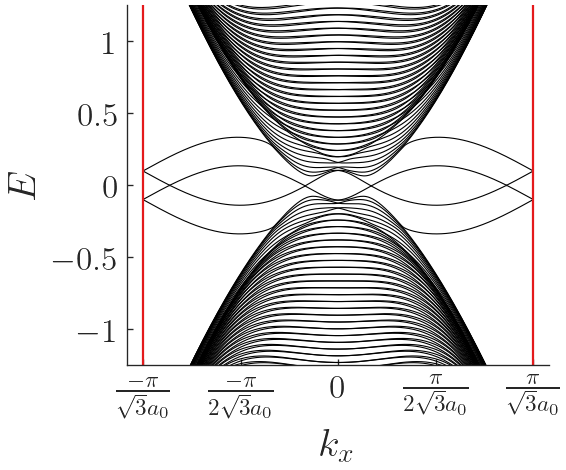}
         \caption{$\xi=2.0$}
         \label{Figk3c}
     \end{subfigure}
\begin{subfigure}{0.49\columnwidth}
         \centering
         \includegraphics[width=\columnwidth]{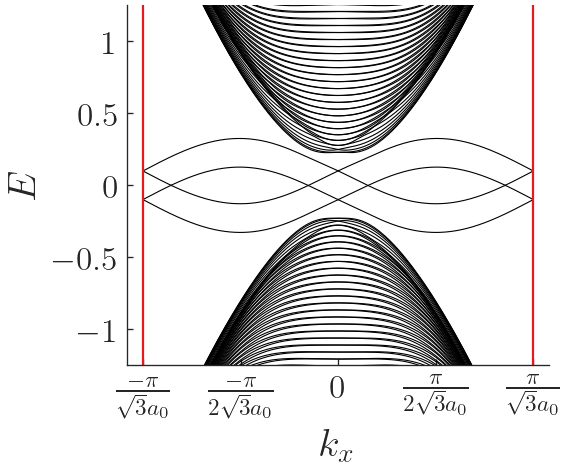}
         \caption{$\xi=2.2$}
         \label{Figk3d}
     \end{subfigure}
\caption{The bandstructure of a finite zig-zag ribbon like configuration for the deformed Kane Mele model is shown. In (a), (b) distinct helical edge states are seen traversing the band gap as long as $\xi<2$. (c), (d) show the scenario for $\xi=2$ and beyond. We observe that beyond the critical value of deformation, the edge states do not traverse the band gap and hence carry no topological significance anymore. This indicates at a destruction of the QSH phase beyond the critical point.}
\label{Figk3}
\end{figure}
\begin{figure}
    \includegraphics[width=0.5\columnwidth]{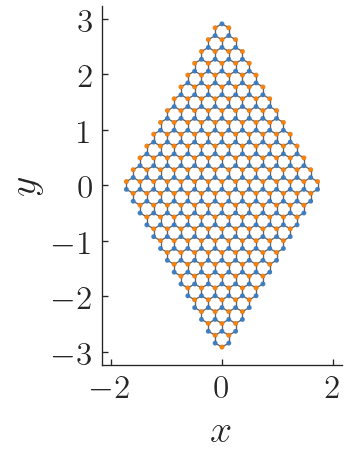}
    \caption{A rhombic supercell constructed using the honeycomb lattice that is used in the calculations of the second order topological phase.}
\label{Figk4}
\end{figure}
\begin{figure}
\centering
\begin{subfigure}{0.85\columnwidth}
         \centering
         \includegraphics[width=\columnwidth]{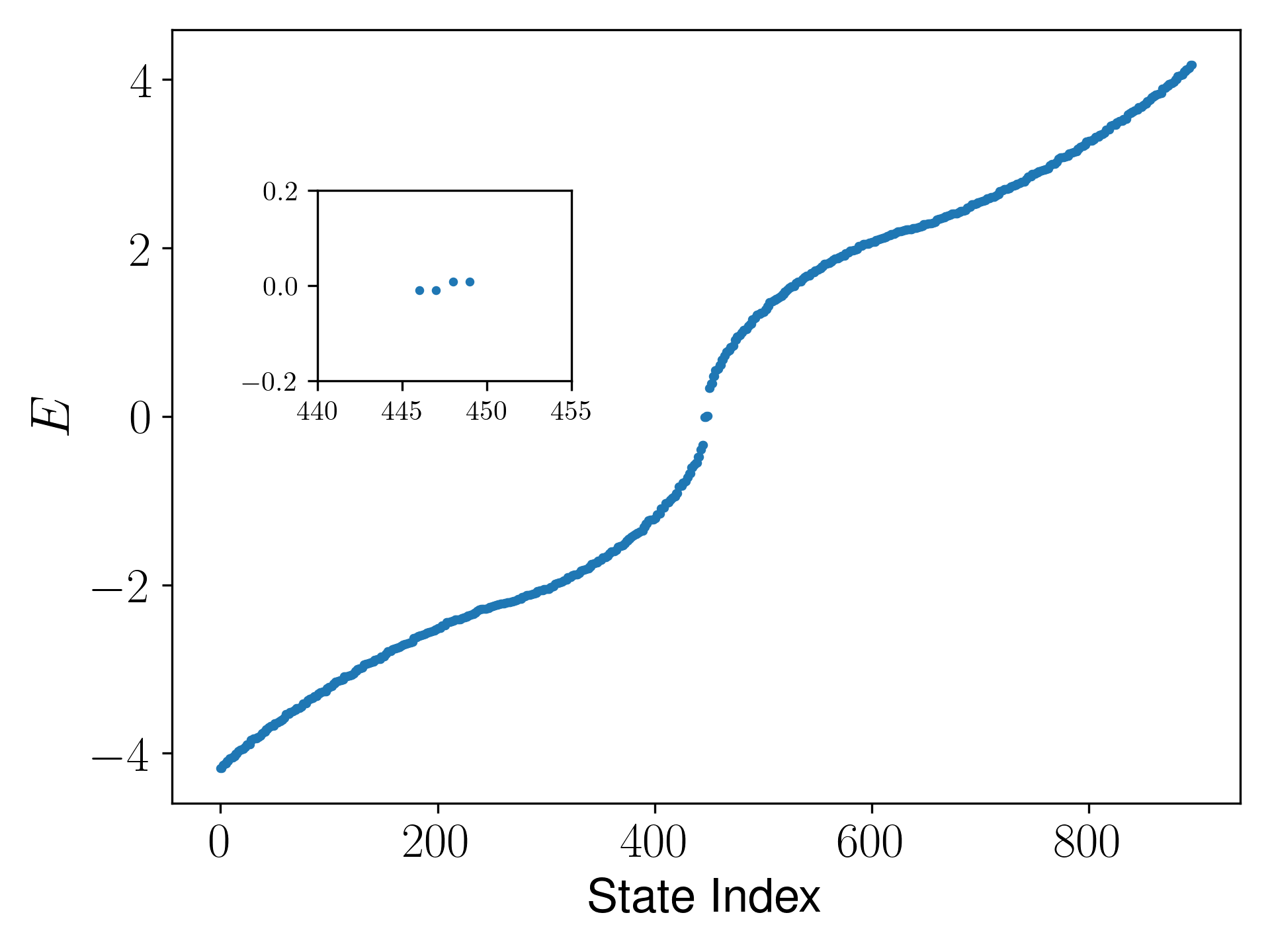}
         \caption{$\lambda_v=0.0$}
         \label{Figk5a}
     \end{subfigure}
\begin{subfigure}{0.85\columnwidth}
         \centering
         \includegraphics[width=\columnwidth]{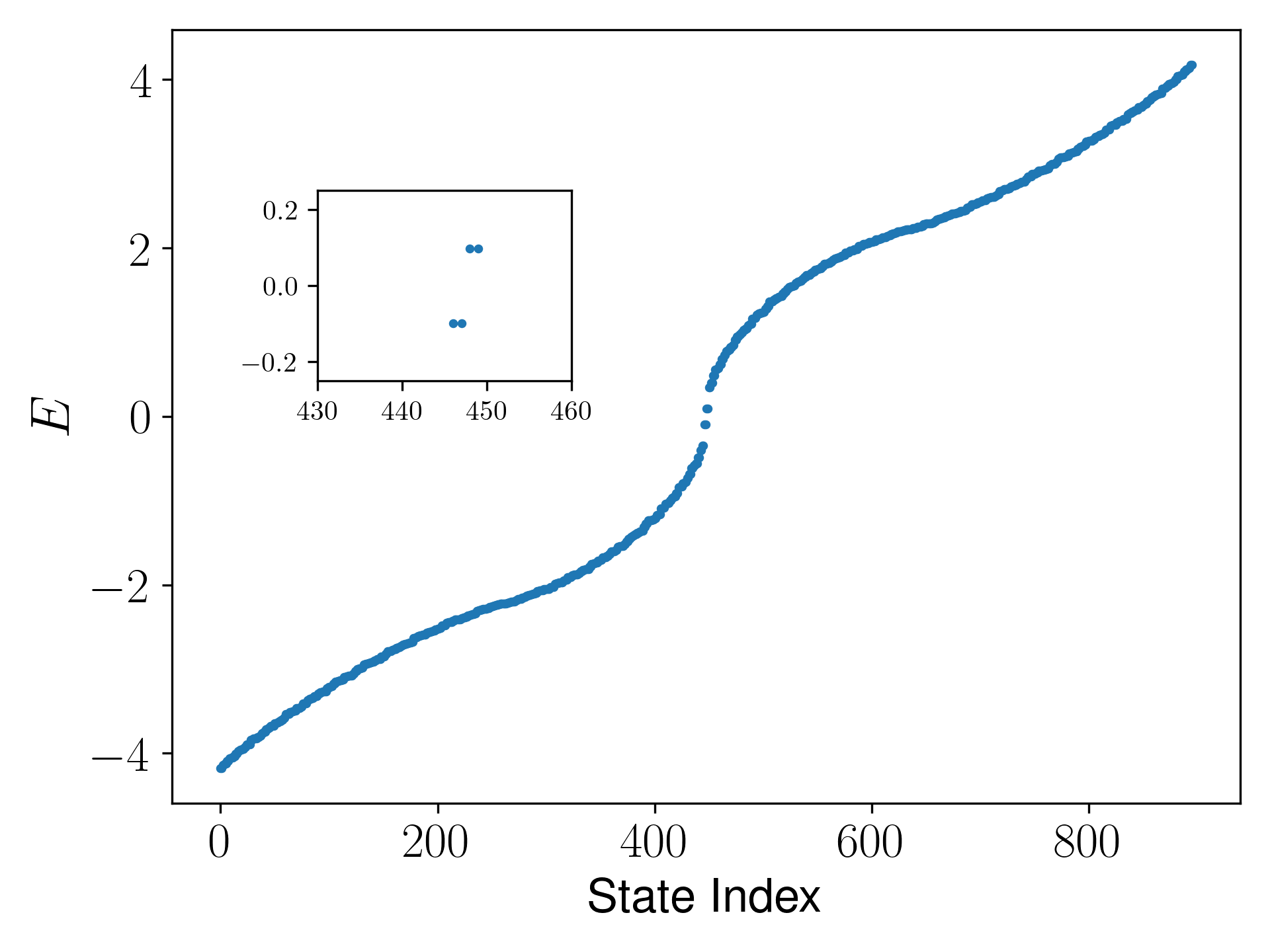}
         \caption{$\lambda_v=0.1t$}
         \label{Figk5b}
     \end{subfigure}
\caption{The real space energy eigenspectra is plotted for two different values of the Semenoff mass $\lambda_v$. The value of the deformation parameter is fixed at $\xi=2.2$. Furthermore, the values of $\lambda_{so}$ and $\lambda_R$ are fixed at $0.06t$ and $0.05t$ respectively. (a) The presence of energy eigenstates, separated from the bulk and pinned at zero energy is seen in the spectrum. These states are four fold degenerate as long as the semenoff mass $\lambda_v$ is kept $0$. (b) For $\lambda_v=0.1t$, the inversion symmetry in the system is broken and the states shift from zero energy. The inset in (a) and (b) zoom in the vicinity of the region $E=0$ and resolves the four states near zero energy.}
\label{Figk5}
\end{figure}
\begin{figure}
    \includegraphics[width=0.49\columnwidth]{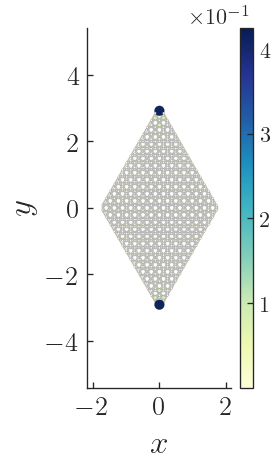}
    \caption{Real space probability density of one of the four degenerate zero energy corner states is shown. The above plot is obtained for $\lambda_{so}=0.06t$, $\lambda_R=0.05t$, $\lambda_v=0$ and $\xi=2.2$. It is seen that the probability densities appear at the vertices.}
\label{Figk6}
\end{figure}
The vectors connecting the nearest neighbours are given by $\vec\delta_1=a_0(0,1)$, $\vec\delta_2=a_0(-\frac{\sqrt{3}}{2},-\frac{1}{2})$, $\vec\delta_3=a_0(\frac{\sqrt{3}}{2},-\frac{1}{2})$ where $a_0$ is the nearest neighbour distance. The lattice vectors are given by $\vec{a}_1=\vec{\delta}_1-\vec{\delta}_2$ and $\vec{a}_2=\vec{\delta}_1-\vec{\delta}_3$. The hexagonal lattice has two sublattices denoted by $A$ and $B$. In our model, the NN hopping along the direction $\hat \delta_1$ is assumed to be $t_1$, while it is given by $t$ in the directions $\hat\delta_2$ and $\hat\delta_3$. We tune the bandstructure as a function of the deformation parameter $\xi=\frac{t_1}{t}$ and observe the behaviour of the boundary states. The tight binding Hamiltonian for the real space Kane Mele model is given as,
\begin{align}
    \begin{split}
        H&=\sum_{\langle i,j \rangle}t_{ij}c_i^\dagger c_j + i\lambda_{so}\sum_{\langle\langle i,j \rangle\rangle}\nu_{ij}c_i^\dagger\sigma_z c_j \\&+ i\lambda_{R}\sum_{\langle i,j \rangle}c_i^\dagger(\boldsymbol\sigma\times\mathbf{\hat d_{ij}})_zc_j+\sum_{i}\lambda_vc_i^\dagger c_i
    \end{split}
\end{align}
where $c_i$ ($c_i^\dagger$) represent annihilation (creation) operators at lattice site $i$. Here $t_{ij}$ is the NN hopping amplitude which is equal to $t_1$ when the hopping occurs along the direction $\vec{\delta}_1$ and is equal to $t$ along $\vec{\delta}_2$ and $\vec{\delta}_3$. The second term is a spin-orbit coupling (SOC) term where $\lambda_{so}$ corresponds to the intrinsic SOC amplitude which is a key ingredient in the formation of the QSH phase. $\nu_{ij}=1$($-1$) if the electron takes a left(right) turn while moving from site $j$ to site $i$. The third term corresponds to Rashba SOC with $\lambda_{R}$ as the coupling strength. The conservation of the $z$ component of spin that is $\sigma_z$ is violated in presence of $\lambda_R$. $\mathbf{\hat d_{ij}}$ corresponds to the nearest neighbor vector connecting site $j$ to site $i$. Finally the fourth term denotes the onsite sublattice potential where $\lambda_v$ assumes a positive amplitude (say $m_s$) for sublattice $A$ and negative (say $-m_s$) for sublattice $B$. It is known that the QSH phase survives in the original Kane-Mele as long as $\lambda_v<3\sqrt{3}\lambda_{so}$ \cite{KM}. Fourier transformation of the real space Hamiltonian gives us the tight binding Hamiltonian in the momentum space,
\begin{align}
\begin{split}
H(\mathbf k)=\begin{pmatrix}
\gamma(\mathbf k) + m_s&\eta(\mathbf k)&0&\rho(\mathbf k)\\
\eta^*(\mathbf k)&-\gamma(\mathbf k)-m_s&-\rho(-\mathbf k)&0\\
0&-\rho^*(-\mathbf k)&-\gamma(\mathbf k)+m_s&\eta(\mathbf k)\\
\rho^*(\mathbf k)&0&\eta^*(\mathbf k)&\gamma(\mathbf k)-m_s
\end{pmatrix}
\label{2}
\end{split}
\end{align}
where
\begin{subequations}
\begin{alignat}{2}
\eta(\mathbf k)&=t_1e^{-ik_ya}+2te^{\frac{ik_ya}{2}}\text{cos}\frac{\sqrt{3}k_xa}{2}\\
\gamma(\mathbf k)&=2\lambda_{so}\Big{[}2\text{sin}\frac{\sqrt{3}k_xa}{2}\text{cos}\frac{3k_ya}{2}-\text{sin}\sqrt{3}k_xa\Big{]}\\
\rho(\mathbf k)&=i\lambda_R\Big{[}e^{-ik_ya}+e^{\frac{ik_ya}{2}}2\text{cos}[\frac{\sqrt{3}k_xa}{2}+\frac{\pi}{3}\Big{]}
\end{alignat}
\end{subequations}
The bulk bandstructure calculated using Eq. \ref{2} shows band extrema at the \textbf{K}($\frac{-2\pi}{3\sqrt3 a_0}, \frac{2\pi}{3a_0}$) and \textbf{K$'$}($\frac{2\pi}{3\sqrt3 a_0},\frac{2\pi}{3a_0}$) points for $\xi=\frac{t_1}{t}=1$, as seen in Fig. \ref{Figk1}. In this case the amplitude of the Rashba SOC and the onsite sublattice potential are kept zero resulting in the spin-$\uparrow$ and spin-$\downarrow$ bands to be degenerate. It is seen that as the band is slowly deformed, the extrema slowly shift towards each other, finally converging at the $M$ point of the BZ for $\xi=2$. The gap closing transition at $\xi=2$ destroys the QSH phase and renders the system trivial from the perspective of first order topology. For non-zero values of the onsite potential $\lambda_v$ and $\lambda_{so}$, the degeneracy of the bands is lifted as seen in Fig. \ref{Figk2}. However, the general behavior of the spectral properties with respect to the deformation parameter remains the same.\par Next, in order to study the behavior of the edge modes pertaining to the QSH phase, we plot the energy bandstructure of a zig-zag ribbon-like configuration with periodic boundary condition (PBC) along the direction $\hat a_1-\hat a_2$ and open boundary condition (OBC) along the direction $\hat a_1$. The presence of PBC along the $x$ direction (which is the same as the direction $\hat a_1-\hat a_2$) enables us to Fourier transform the Hamiltonian along $x$-direction and thus plot the dispersion of this finite ribbon as a function of $k_x$. Distinct edge modes are seen traversing the band gap as a function of $k_x$, in the region $1<\xi<2$, as shown in Fig. \ref{Figk3}. Evidently, these are conducting eigenstates confined to the edges of the system. At $\xi=2$, the closure of the bulk band gap causes the first order topological phase to disappear and the conducting edge states get trivialised beyond this critical point.\par To investigate the topology of the phase beyond the critical point ($\xi=2$), we carefully construct a rhombic supercell taking into the account that the system possesses a mirror symmetry $M_x$. A schematic representation of this supercell is shown in Fig. \ref{Figk4}. The real space energy eigenspectra is evaluated which shows the presence of four distinct zero energy modes. In presence of a non-zero onsite potential $\lambda_v$, the in-gap modes shift from zero energy as shown in Fig. \ref{Figk5}. The real space probability distribution of the zero energy states show that they are confined at the two mirror invariant corners of the rhombic supercell (Fig. \ref{Figk6}).
\section{\label{sec:level2}Topological invariants}
The QSH phase seen in the regime $\xi<2$, is a $\mathbb{Z}_2$ topological phase which has a zero Hall conductivity but a non-zero spin Hall conductivity. If the $z$-component of spin, that is $\sigma_z$, is conserved (for the case where $\lambda_R=0$), the spin Hall conductivity is also quantized. The $\mathbb{Z}_2$ invariant in such a case is given by \cite{Z21, Z22},
\begin{align}
\begin{split}
\nu = (C^\uparrow-C^\downarrow)/2
\end{split}
\end{align}
where $C^\uparrow$ ($C^\downarrow$) refers to the spin-$\uparrow$ (spin-$\downarrow$) Chern numbers. However, in presence of a Rashba SOC, the $z$-component of spin that is $\sigma_z$ is not conserved and hence this form of the $\mathbb{Z}_2$ invariant is no longer valid. However, a quantized $\mathbb{Z}_2$ invariant pertaining to a quantum spin Hall phase still persists. In our work we study the hybrid Wannier charge centers as a function of the deformation parameter $\xi$ to follow the fate of the QSH phase in the presence of $\sigma_z$ non-conserving terms. In this context, Wannier charge center refers to the center of charge in a unit cell.
\begin{figure}
\centering
\begin{subfigure}{0.75\columnwidth}
         \centering
         \includegraphics[width=\columnwidth]{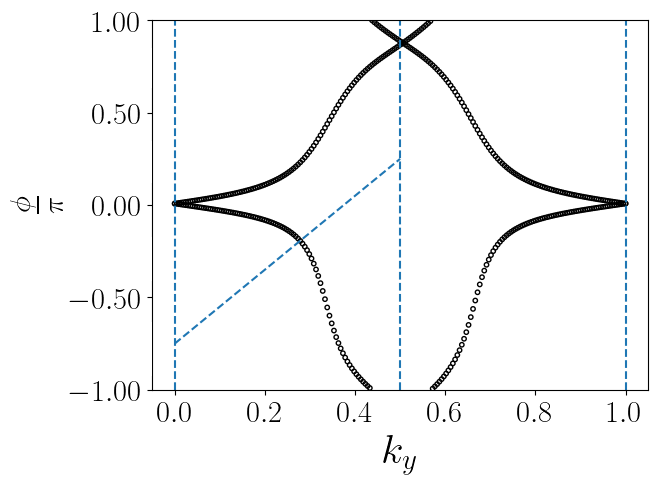}
         \caption{$\xi=1.0$}
         \label{Figk7a}
     \end{subfigure}
\begin{subfigure}{0.75\columnwidth}
         \centering
         \includegraphics[width=\columnwidth]{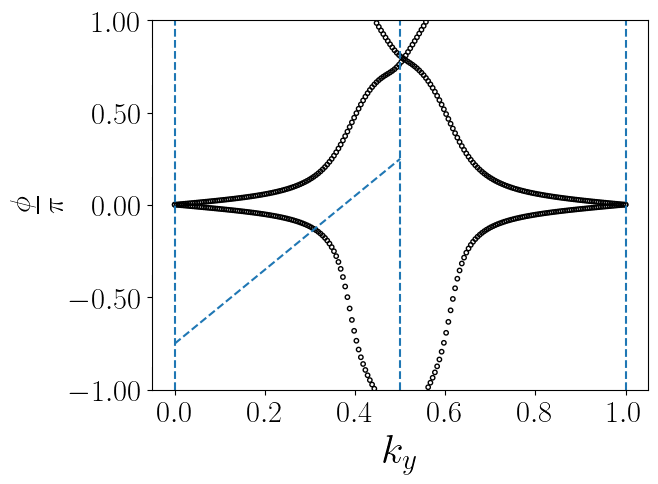}
         \caption{$\xi=1.5$}
         \label{Figk7b}
     \end{subfigure}
\caption{Evolution of the Wannier charge center along $x$-direction as a function of $k_y$ is shown for different values of the deformation parameter $\xi$ in the QSH phase. Clearly, each WCC undergoes a winding as the value of $k_y$ is evolved. The dotted line traversing half the BZ is an indicator of the $\mathbb{Z}_2$ invariant pertaining to the phase of the system. The number of intersections of the dotted line with the WCC curve denotes the $\mathbb{Z}_2$ invariant of the phase.  All the calculations have been done for $\lambda_{so}=0.06t$, $\lambda_R=0.05t$ and $\lambda_v=0.1t$.}
\label{Figk7}
     \end{figure}
\begin{figure}
\centering
\begin{subfigure}{0.75\columnwidth}
         \centering
         \includegraphics[width=\columnwidth]{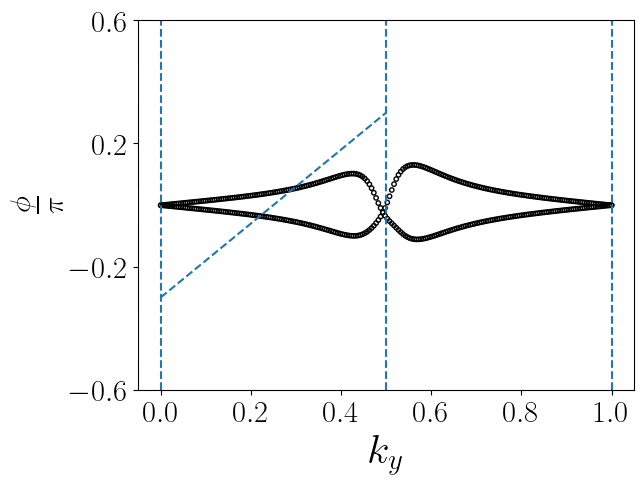}
         \caption{$\xi=2.2$}
         \label{Figk7.5a}
     \end{subfigure}
\begin{subfigure}{0.75\columnwidth}
         \centering
         \includegraphics[width=\columnwidth]{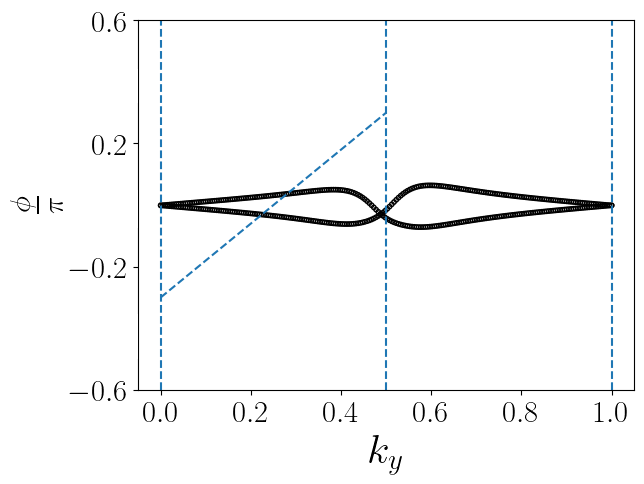}
         \caption{$\xi=2.5$}
         \label{Figk7.5b}
     \end{subfigure}
\caption{Evolution of the Wannier charge center along $x$-direction as a function of $k_y$ is shown for different values of $\xi$ in the HOTI phase. The two WCC undergo no winding here as a function of $k_y$. Furthermore, the number of intersections of the dotted line traversing half the BZ, with the two WCC is even. This indicates a trivial $\mathbb{Z}_2$ invariant. All the calculations have been done for $\lambda_{so}=0.06t$, $\lambda_R=0.05t$ and $\lambda_v=0.1t$.}
\label{Figk7.5}
\end{figure}
\begin{figure}
\centering
\begin{subfigure}{0.49\columnwidth}
         \centering
         \includegraphics[width=\columnwidth]{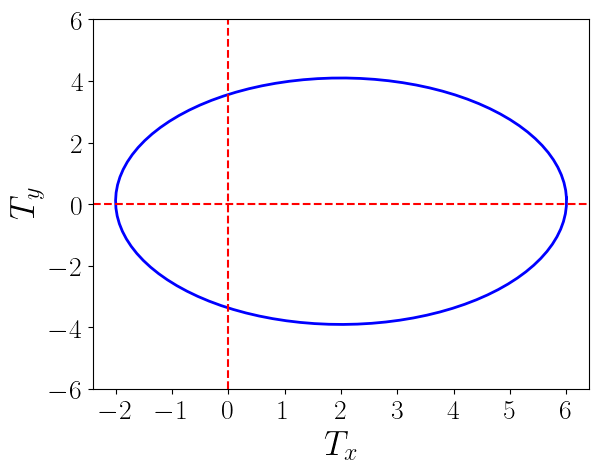}
         \caption{$\xi=1.0$}
         \label{Figk8a}
     \end{subfigure}
\begin{subfigure}{0.49\columnwidth}
         \centering
         \includegraphics[width=\columnwidth]{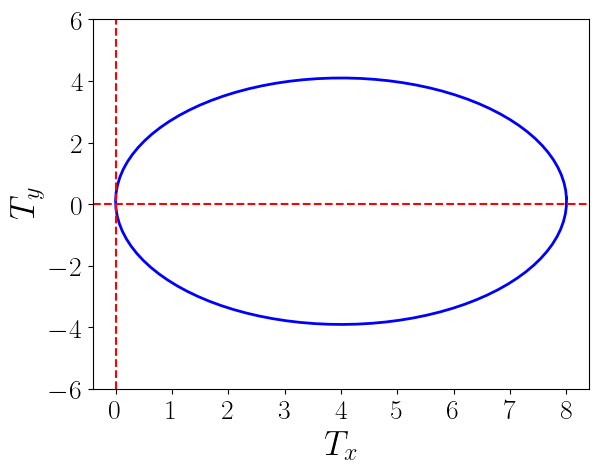}
         \caption{$\xi=2.0$}
         \label{Figk8b}
     \end{subfigure}
\begin{subfigure}{0.49\columnwidth}
         \centering
         \includegraphics[width=\columnwidth]{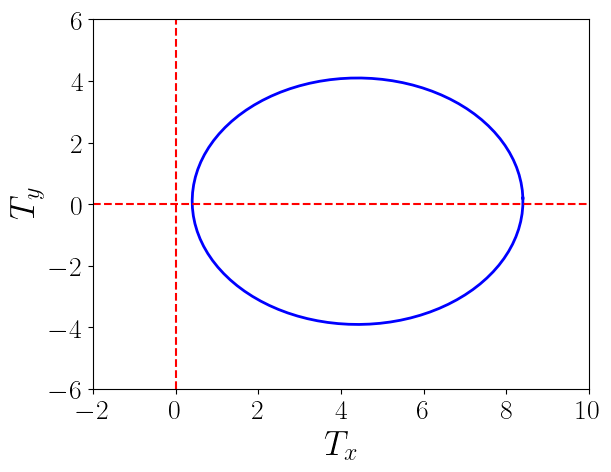}
         \caption{$\xi=2.2$}
         \label{Figk8c}
     \end{subfigure}
\begin{subfigure}{0.49\columnwidth}
         \centering
         \includegraphics[width=\columnwidth]{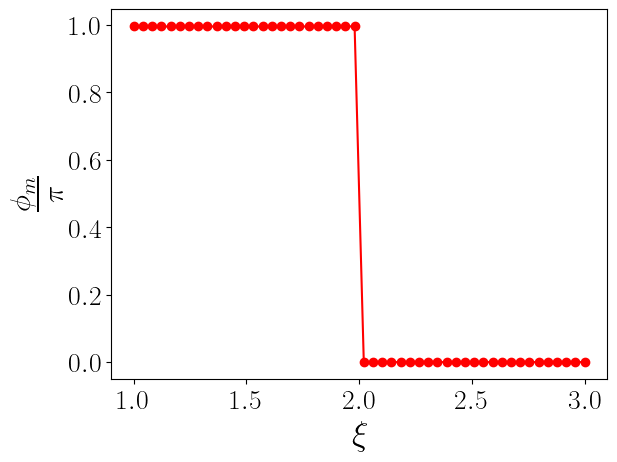}
         \caption{}
         \label{Figk8d}
     \end{subfigure}
\caption{The winding (or the absence of it) of the origin in the $T_x-T_y$ plane of the positive mirror subspace is shown for several values of the deformation parameter $\xi$. (a) For $\xi=1$ it seen that the origin lies within the area enclosed by the curve. (b) $\xi=2$ clearly represents a phase transition (critical or gap closing) point. (c) For $\xi>2$, the origin is no longer wound. Due to the chiral symmetry of the mirror resolved Hamiltonian $H^\pm$, this winding directly corresponds to the Berry phase acquired by the corresponding eigenstate over a complete BZ. This winding number is plotted in (d) for a range of the deformation parameter $\xi$. It is seen that the 1D polarization or the Berry phase is incapable of characterizing the second order topological phase. The values of the parameters have been kept fixed at $\lambda_{so}=0.06t$, $\lambda_R=0.05t$ and $\lambda_v=0$.}
\label{Figk8}
\end{figure}
\begin{figure}
\centering
\begin{subfigure}{0.70\columnwidth}
         \centering
         \includegraphics[width=\columnwidth]{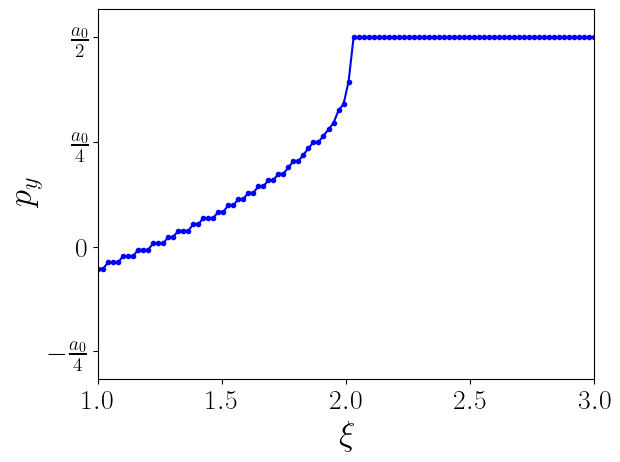}
         \caption{}
         \label{Figk9a}
     \end{subfigure}
\begin{subfigure}{0.70\columnwidth}
         \centering
         \includegraphics[width=\columnwidth]{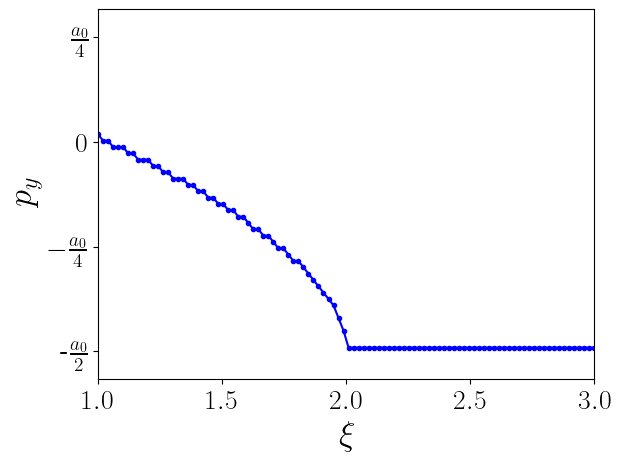}
         \caption{}
         \label{Figk9b}
     \end{subfigure}
\caption{The bulk polarization $p_y$ has been plotted over a range of $\xi$. (a) For the spin-$\uparrow$ eigenstate, the 2D or bulk polarization acquires a quantized value of $\frac{a_0}{2}$ beyond $\xi>2$, that is in the SOTI phase. (b) For the spin-$\downarrow$ phase, the bulk polarization acquires a value of $-\frac{a_0}{2}$. The quantized value of bulk polarization indicates the formation of an \textit{obstructed atomic insulator} where the center of charge does not coincide with the real lattice points. This gives rise to the formation of the SOTI phase.}
\label{Figk9}
\end{figure}
Mathematically they represent the expectation value of position operator for a basis represented by Wannier functions which are a set of orthogonal functions indexed by a lattice position say $\mathbf R$ and maximally localized about that point with respect to all relevant spatial dimensions. The Wannier functions are represented as \cite{WF},
\begin{align}
\begin{split}
|\mathbf R, n\rangle=\frac{V}{{(2\pi)}^D}\int \text d^Dk e^{-i\mathbf k.\mathbf R}|\psi_{n\mathbf k}\rangle
\end{split}
\end{align}
where $|\psi_{n\mathbf k}\rangle$ represents the Bloch wave function and $D$ corresponds to the dimensionality of the $k$-space. $V$ is the real space primitive cell volume. Hybrid Wannier functions, on the other hand, refer to wave functions which are localized along one spatial dimension (say, $x$) while being delocalized along the other dimensions (say, $y$ and $z$) and can be written as,
\begin{align}
    \begin{split}
        |R_x, k_y, k_z, n\rangle=\frac{1}{2\pi}\int_{-\pi}^{\pi}\text{d}k_xe^{-iR_xk_x}|\psi_{n\mathbf{k}}\rangle
    \end{split}
\end{align}
Expectation value of the position operator (say, $\hat{X}$) with respect to the hybrid Wannier function, gives us the hybrid Wannier charge center. Mathematically, this is represented as,
\begin{align}
\begin{split}
\bar x_n(k_y, k_z) = \langle R_x, k_y, k_z, n|\hat X|R_x, k_y, k_z, n\rangle
\end{split}
\end{align}

The hybrid WCC being proportional to the Berry phase captures the topological details of the system efficiently and is given as \cite{WCC_1, WCC_2},
\begin{align}
    \begin{split}
        \phi_n(k_y)=\int_{0}^{2\pi}A_n(k_x,k_y)dk_x
    \end{split}
\end{align}
Here $\mathbf{A_n}(k_x, k_y)=-i\langle u_{nk}|\nabla_k|u_{nk}\rangle$ is known as the Berry connection, where $n$ is the band index and $|u_{nk}\rangle$ corresponds to the periodic part of the Bloch wavefunction. Here, we calculate the WCC along the $x$-direction and study its evolution as a function of the momentum in the $y$-direction that is $k_y$ (since the model we study is 2D and lies on the $x-y$ plane). The $\mathbb{Z}_2$ invariant is now defined as the number of individual hybrid WCC crossed by a line traversing half the BZ, modulo $2$ \cite{WCC}. If the line cuts through odd (even) number of hybrid WCC while traversing half the BZ, the $\mathbb{Z}_2$ invariant is non-trivial (trivial). We observe in Fig. \ref{Figk7} that the $\mathbb{Z}_2$ invariant remains non-trivial as long as $\xi<2$. Beyond this point, the evolution of the hybrid WCC is changed and the system no longer remains in the QSH phase, as shown in Fig. \ref{Figk7.5}.\par
Next we focus on the crystalline symmetries of the deformed Kane-Mele Hamiltonian. The deformed Kane-Mele model possesses a mirror symmetry $M_x$ given by $s_x\otimes \mathbb I$, where $s_x$ and $\mathbb I$ act on the spin and the sublattice degrees of freedom respectively. Here $\mathbb{I}$ corresponds to identity and $s_x$ corresponds to the $x$ component of the Pauli matrices. The mirror symmetry decouples the Hamiltonian into two subspaces given by the positive and negative mirror eigenvalues. We put $k_x=0$ and decouple the Hamiltonian $H(0, k_y)$ into two parts denoted by $H^\pm$ corresponding to the positive and the negative values of the mirror symmetry operator $M_x$. The action of the mirror symmetry operator on the Hamiltonian is given as follows,
\begin{align}
M_xH(k_x, k_y)M_x^{-1}=H(-k_x, k_y)
\end{align}
Thus, on putting $k_x=0$, the mirror operator $M_x$ can be used to decouple the original Hamiltonian into $H^\pm$ which is given as,
\begin{align}
    \begin{split}
        H^\pm(k_y)=T^\pm_x(k_y)\sigma_x+T_y^\pm(k_y)\sigma_y
    \end{split}
\end{align}
where,
\begin{subequations}
\begin{alignat}{2}
T_x^\pm(k_y)=\pm2\lambda_R\text{sin}\frac{3ak_y}{2}+4t\text{cos}\frac{3ak_y}{2}+2t_1\\
T_y^\pm(k_y)=\pm2\lambda_R\Big{[}\text{cos}\frac{3ak_y}{2}+1\Big{]}-4t\text{sin}\frac{3ak_y}{2}
\end{alignat}
\end{subequations}
On studying the evolution of $T_x$ and $T_y$ over a complete path in the BZ ($\Gamma\rightarrow M \rightarrow\Gamma$), we see that the winding number is 1 (that is the origin of the $T_x-T_y$ plane is enclosed) only when the deformation parameter $\xi$ remains less than $2$. The origin lies outside the enclosed area as soon as the QSH phase is destroyed and the second order topological phase is reached, as shown in Fig. \ref{Figk8a}, \ref{Figk8b}, \ref{Figk8c}. This implies that the Berry phase acquired by the ground state of either the positive or negative subspace of the mirror resolved Hamiltonian, is another alternate bulk property that correctly captures the QSH phase. However, it is trivial in the second order topological regime. This has been shown in Fig. \ref{Figk8d}, where $\phi_m$ represents the Berry phase acquired by the ground state of the effective 1D mirror resolved Hamiltonian $H^+(k_y)$ over a complete 1D path in the BZ, and is given by,
\begin{align}
    \phi_m = -i\int_{\Gamma\rightarrow M\rightarrow \Gamma}\text{d}k_y\langle u_{nk_y}|\nabla_{k_y}|u_{nk_y}\rangle
\end{align}
$|u_{nk_y}\rangle$ corresponds to the periodic part of the Bloch wavefunction $|\psi_{nk_y}\rangle$ belonging to the band $n$. The negative mirror subspace given by $H^-(k_y)$ shows a similar behavior. Thus it is implied that both the evolution of the WCC and the 1D polarization corresponding to the effective Hamiltonian $H^\pm(k_y)$ are incapable of capturing any essence of the second order topological phase (that is the regime beyond $\xi>2$) whereas they accurately characterize the first order QSH phase.\par
To characterize the second order topological states of the Kane Mele model beyond $\xi>2$, we resort to spin resolved bulk polarization. Keeping the value of $\lambda_R=0$, so that the $z$-component of spin is conserved, we calculate bulk polarization for the two different spin sectors which is given by \cite{EZAWA1},
\begin{align}
\begin{split}
\label{Eq 6}
 p^s_\alpha={\bar r_\alpha} = \langle w^s_n|r_\alpha|w^s_n\rangle=\frac{i}{S}\int_{BZ} d^{d}k \langle u^s_{nk}|\frac{\partial}{\partial k_{r_\alpha}}|u^s_{nk}\rangle
\end{split}
\end{align}
where $|w_n^s\rangle=|\mathbf{0},n\rangle_s$ is the Wannier function corresponding to the $n^{th}$ band and $p_\alpha^s$ refers to the value of bulk polarization in the direction $\alpha$ for the spin component $s$ ($\uparrow,\downarrow$). $S$ corresponds to the total area of the honeycomb BZ and is taken as ${8\pi^2}\mathbin{/}{3\sqrt{3}a_0^2}$. As shown in Fig. \ref{Figk9} we observe that the bulk polarization $p_y$ has a quantized value of $\frac{a_0}{2}$ for the spin-$\uparrow$ and a value of $-\frac{a_0}{2}$ for the spin-$\downarrow$ component for  the regime $\xi>2$. For $1<\xi<2$, that is in the QSH phase $p_y$ bears no quantized value. Furthermore it is seen that the value of $p_x$ is uniformly zero both above and below the critical point $\xi=2$. Thus, we establish that the second order topological phase of the band deformed Kane Mele model is an \textit{obstructed atomic insulator} phase where the center of charge in a unit cell is displaced from the actual lattice point in real space and lies between two consecutive sites. The displacement of the center of charge results in fractional charge accumulation at two specific corners of the rhombic supercell, thus exhibiting second order topology in the form of localized corner modes.
\section{\label{sec:level2}Conclusion}
We study a prototypical quantum spin Hall system that shows two topological phases of different orders, brought about by band deformation. The system under study is the celebrated Kane Mele model on a honeycomb lattice which exhibits the presence of helical edge modes as a signature of the quantum spin Hall phase. We smoothly deform the bandstructure of the Kane Mele model by varying one of the nearest neighbour hopping amplitudes (say $t_1$) of the honeycomb lattice while keeping the other two (say $t$) fixed. It is observed that the system retains the QSH phase as long as $\frac{t_1}{t}<2$. We plot the bandstructure of a zig-zag ribbon like configuration to explicitly show the helical edge states which disappear after the system is deformed beyond $\frac{t_1}{t}=2$. However, beyond this critical point is crossed, the system transcends into a second order topological phase hosting robust second order modes at the two corners of a suitably formed rhombic supercell. We study bulk properties like evolution of the hybrid WCC, mirror winding number and spin resolved bulk polarization to characterize and study the evolution of the different topological phases. The evolution of the hybrid WCC shows a stark contrast in the first order and the second order topological phase. The nature of the evolution establishes that the $\mathbb{Z}_2$ invariant is non-trivial in the QSH phase, while being trivial in the HOTI phase. The mirror winding number shows a similar behaviour where it is non-trivial only in the QSH phase. Finally, to decipher the origin of the second order topological phase we calculate spin resolved bulk polarization which depicts the center of charge in a unit cell. We observe that for $\xi>2$, the value of $|p_y|$ is quantized for both the spin sectors at $\frac{a_0}{2}$, while it is not quantized for $\xi<2$. $p_x$ on the other hand is uniformly zero everywhere. This quantization of the value of $p_y$ indicates a displacement of center of charge with respect to the real space lattice site and causes the appearance of fractional charge excess at the corners of the rhombic supercell. The bulk polarization $p_y$ for the spin-$\uparrow$ sector is $\frac{a_0}{2}$ while it is $-\frac{a_0}{2}$ for the spin-$\downarrow$ sector. Herein lies the appearance of a second order topological phase beyond $\xi>2$ which is an obstructed atomic insulator. 
\bibliographystyle{ieeetr}
\bibliography{ref.bib}
\end{document}